\documentclass[12pt]{article}

\usepackage{scicite}

\usepackage{times}
\usepackage{graphicx}

\newenvironment{sciabstract}{%
\begin{quote} \bf}
{\end{quote}}

\newcounter{lastnote}
\newenvironment{scilastnote}{%
\setcounter{lastnote}{\value{enumiv}}%
\addtocounter{lastnote}{+1}%
\begin{list}%
{\arabic{lastnote}.}
{\setlength{\leftmargin}{.22in}}
{\setlength{\labelsep}{.5em}}}
{\end{list}}

\title{Particle Acceleration on Megaparsec Scales in a Merging Galaxy Cluster\footnote{This is the author's version of the work. It is posted here by permission of the AAAS for personal use, not for redistribution. The definitive version was published in Science, {Volume 330, 15 October 2010}.}}

\author
{Reinout J. van Weeren,$^{1\ast}$ Huub J. A. R\"ottgering,$^{1}$ Marcus Br\"uggen$^{2}$, Matthias Hoeft$^{3}$\\
\\
\normalsize{$^{1}$Leiden Observatory, Leiden University,}\\
\normalsize{P.O. Box 9513, NL-2300 RA, Leiden, The Netherlands}\\
\normalsize{$^{2}$Jacobs University Bremen, P.O. Box 750561, 28725 Bremen, Germany}\\
\normalsize{$^{3}$Th\"uringer Landessternwarte, Sternwarte 5, 07778 Tautenburg, Germany}\\
\\
\normalsize{\vspace{-0.05cm}$^\ast$To whom correspondence should be addressed; E-mail:  rvweeren@strw.leidenuniv.nl}
}

\date{}

\begin{document} 


\baselineskip24pt


\maketitle


\begin{sciabstract} 
Galaxy clusters form through a sequence of mergers of smaller   
galaxy clusters and groups. 
Models of diffusive shock acceleration (DSA) suggest that in shocks 
that occur during cluster mergers, particles are accelerated to relativistic 
energies, similar to supernova remnants. 
Together 
with magnetic fields these particles emit synchrotron radiation and may 
form so-called radio relics.  
Here we report the detection of a radio relic for 
which we find highly aligned magnetic fields, a strong spectral 
index gradient, and a narrow relic width, giving a measure 
of the magnetic field in an unexplored site of the universe. 
Our observations prove that DSA also operates on scales much 
larger than in supernova remnants and that shocks in galaxy 
clusters are capable of producing extremely energetic cosmic rays. 

\end{sciabstract}

In the universe structure forms hierarchically with smaller structures 
merging to form bigger ones. On 
the largest scales, clusters of galaxies merge releasing energies of 
the order of $10^{64}$~erg on timescales of 1--2~Gyr \cite{1999ApJ...518..603R,2008MNRAS.391.1511H}. 
During such merger events, large-scale shock waves with moderate Mach numbers of 1--5, should be created. 
In such shocks, DSA is expected to accelerate electrons to high energies; in the presence of a magnetic field, 
these particles are expected to form large regions emitting synchrotron 
radiation at radio wavelengths \cite{2001ApJ...562..233M,2008MNRAS.391.1511H,2008MNRAS.385.1211P }. 
The accelerated particles at the shock front have 
a power-law energy distribution which directly translates into an 
integrated power-law radio spectrum (flux $\propto \nu^{\alpha}$, 
with $\alpha$ the spectral index and $\nu$ the frequency). 
The slope of the particle distribution ($s$) in the linear test particle regime, and thus the radio spectral index 
($\alpha=(1-s)/2$), only depends on the compression ratio ($r$) of the shock \cite{1983RPPh...46..973D,1987PhR...154....1B}, with $s=(r+2)/(r-1)$. 
At the shock front,  
the intracluster medium (ICM) is compressed such that magnetic fields  
align parallel to the shock front \cite{1998A&A...332..395E}. 
These ordered and aligned magnetic fields cause the radio emission to be highly polarized. 
Synchrotron and inverse Compton (IC) losses cool the radio plasma behind the shock, creating a 
strong spectral index gradient in the direction towards the cluster center. 
It has been suggested that such synchrotron emitting regions from shocks can 
be identified with radio relics \cite{1998A&A...332..395E,2001ApJ...562..233M}. 
These are elongated radio sources located mostly in 
the outskirts of massive merging galaxy 
clusters \cite{2005AdSpR..36..729F,2006Sci...314..791B,2006AJ....131.2900C,2009A&A...503..707B,2009A&A...494..429B,2009A&A...505..991V,2009A&A...506.1083V}.

Here we present the detection of a 2~Mpc radio relic (\verb Figs. 1,~2)  
located in the northern outskirts of the galaxy cluster CIZA~J2242.8+5301 ($z=0.1921$). 
This X-ray luminous cluster \cite{2007ApJ...662..224K} ($L_{X} = 6.8 \times 10^{44}$~ergs$^{-1}$, 
between 0.1 and 2.4~keV) shows a disturbed elongated morphology in ROSAT X-ray images \cite{1999A&A...349..389V}, 
indicative of an undergoing major merger event. 
The relic is located at a distance of 1.5~Mpc from the cluster center. 
Unlike other known radio relics, the northern relic is 
extremely narrow with a width of 55~kpc. 
Deep Westerbork Synthesis Radio 
Telescope (WSRT), Giant Metrewave Radio Telescope (GMRT) and 
Very Large Array (VLA) observations (SOM) show a clear unambiguous 
spectral index gradient towards the cluster center (\verb Fig. 3). 
The spectral index, measured over a range of frequencies between 2.3 and 0.61~GHz, steepens from $-0.6$ to $-2.0$ across 
the width of the narrow relic. The gradient is visible over the entire 2~Mpc 
length of the relic, constituting clear evidence for shock 
acceleration and spectral ageing of relativistic electrons 
in an outward moving shock. 
The relic's integrated radio 
spectrum is a single power-law, with $\alpha   = -1.08 \pm 0.05$, 
as predicted \cite{1983RPPh...46..973D,1987PhR...154....1B}. 
The relic is strongly polarized at the 50--60\% level, 
indicating a well ordered magnetic field, and polarization magnetic field vectors 
are aligned with the relic. 
In the southern part of the cluster, located symmetrically 
with respect to the cluster center and the northern relic, there is a second fainter broader relic. The elongated 
radio relics are orientated perpendicular to the major axis of the cluster's elongated X-ray emitting ICM, as expected for 
a binary cluster merger event 
in which the second southern relic traces the opposite shock 
wave \cite{1999ApJ...518..603R}. 
Furthermore, there is a faint halo of diffuse radio emission extending all the way towards the cluster center 
connecting the two radio relics (\verb Fig. 1). 
This emission extends over 3.1~Mpc, making it the largest known diffuse radio source in a cluster to date.

The source cannot be a gravitational lens because it is too large and 
located too far from the cluster center. 
The morphology, spectral index, and association with a cluster 
exclude the possibility of the source being a supernova remnant. 
The source is also not related to the radio AGN 
located at the eastern end of the relic. High-resolution 
observations show this source to be detached from the relic (\verb Fig. 2). 
The spectral and polarization properties are also unlike that 
of any known tailed radio sources \cite{1973A&A....26..413M,1998A&A...331..901S}. 
The power-law radio spectral index, clear spectral index gradient 
and enormous extent, exclude the possibility the source is tracing (compressed) fossil 
radio plasma from a radio source whose jets are now off \cite{2001A&A...366...26E,2002MNRAS.331.1011E}. 
The integrated radio spectra of such fossil sources are very steep ($\alpha < -1.5$) 
and curved, because the radio emitting plasma is old and has undergone synchrotron and IC losses. 
In addition, the shell-like (and not lobe-like) morphology 
does not support the above scenario.

Instead, all the observed properties of the relic perfectly match that of electrons 
accelerated at large-scale shocks via DSA. 
The characteristics of the bright relic provide evidence that (at least some) relics 
are direct tracers of shocks waves, 
and a way to determine the magnetic field strength at the 
location of the shock using similar arguments to those that have been used 
for supernova remnants \cite{2003ApJ...584..758V}.

The configuration of the relic arises naturally for a roughly equal mass head-on binary 
cluster merger, without much substructure, in the plane of the sky with the shock waves seen edge-on. First, 
the polarization fraction of 50\% or larger can only explained 
by an angle of less than 30 degrees between the line-of-sight 
and the shock surface \cite{2006AJ....131.2900C}. Second, because there is evidence for 
spectral ageing across the relic, only part of the width can be caused by projection effects.

The amount of spectral ageing by synchrotron and IC losses 
is determined by the magnetic field strength, $B$, 
the equivalent magnetic field strength of the cosmic 
microwave background (CMB), $B_{\rm{CMB}}$, and the observed frequency. 
The result is a downward spectral curvature resulting in a steeper 
spectral index in the post-shock region (i.e., lower $\alpha$). 
For a relic seen edge-on the downstream luminosity and spectral 
index profiles thus directly reflect the ageing of the relativistic electrons \cite{2005ApJ...627..733M}. 
To first approximation, the width of the relic ($l_{\rm{relic}}$) is 
determined by a characteristic timescale ($t_{\rm{sync}}$) due to spectral ageing, 
and the downstream velocity 
($v_{d}$): $l_{\rm{relic}} \approx t_{sync} \times v_{d}$, 
with $t_{sync} \propto  \left( B^{1/2}/(B^2 + B_{\rm{CMB}}^2) \right) \times \left(1/(\nu(1+z))^{1/2}\right)$. 
Conversely, from the width of the relic and its downstream velocity, 
a direct measurement of the magnetic field at the location of 
the shock can be obtained. Using standard shock jump conditions, 
it is possible to determine the downstream velocity, 
from the Mach number and the downstream plasma temperature.

The spectral index at the front of the relic is $-0.6 \pm 0.05$ which 
gives a Mach number of $4.6_{-0.9}^{+1.3}$ for the 
shock \cite{2009A&A...506.1083V} in the linear regime. Using the $L_{X}-T$ scaling 
relation for clusters \cite{1998ApJ...504...27M} we estimate 
the average temperature of the ICM to be $\sim9$~keV. Behind the shock front 
the temperature is likely to be higher. Temperatures 
in the range between 1.5 and 2.5 times the average value have 
previously been observed \cite{2009ApJ...693L..56M}. The derived Mach number and the advocated 
temperature range, imply downstream velocities between 900 and 1,200~km~s$^{-1}$ (we used an adiabatic exponent of $5/3$). 
For the remainder we will adopt a value of 1,000~km~s$^{-1}$. Using 
the redshift, downstream velocity, spectral index, and characteristic 
synchrotron timescale we have for the width of the relic
\begin{equation}
l_{\rm{relic,~610~MHz}}   \approx 1.2 \times 10^{3} \frac{B^{1/2}}{B^2 + B_{\rm{CMB}}^{2}} \mbox{ } \mbox{ [kpc]~, }
\end{equation}
with $B$ and $B_{\rm{CMB}}$ in units of $\mu$Gauss. 
Because $B_{\rm{CMB}}$ is known, the measurement of $l_{\rm{relic}}$ 
from the radio maps, directly constrains the magnetic field.  
From the 610 MHz image (the image with the best signal to noise ratio and highest angular resolution), 
the relic has a deconvolved width (full width at half maximum) of 55~kpc (\verb Fig. 4). 
Because \verb Eq. 1 has two solutions, the strength 
of the magnetic field is 5 or 1.2~$\mu$Gauss. 
However, projection effects can increase the observed 
width of the relic and affect the derived magnetic field strength. 
Therefore, the true intrinsic width of the relic could be 
smaller, which implies that $B \ge 5$ or $\le 1.2$~$\mu$Gauss (\verb Eq. 1). 
We investigated the effects of projection 
using a curvature radius of 1.5~Mpc, the projected distance from the cluster center. 
Instead of using \verb Eq. 1, we computed full radio profiles \cite{2007MNRAS.375...77H}
for different angles subtended by a spherical shock front 
into the plane of the sky ($\Psi$; the total angle subtended is $2\Psi$ for a shell-like relic).  
The profile for $\Psi = 10$~deg and $B = 5$~$\mu$Gauss agrees best 
with the observations (\verb Fig. 4). For $\Psi = 15$~deg, 
$B$ is 7 or $0.6$~$\mu$Gauss. Values of $\Psi$ 
larger than $\sim15$~deg are ruled out. Lower limits 
placed on the IC emission \cite{2009PASJ...61..339N,2010ApJ...715.1143F} and 
measurements of Faraday rotation \cite{2001ApJ...547L.111C} indicate magnetic 
fields higher than $\sim2$~$\mu$Gauss. We therefore exclude the lower solutions 
for the magnetic field strength and conclude that the magnetic field at the location 
of the bright radio relic is between 5 and 7~$\mu$Gauss.

Magnetic fields within the ICM are notoriously difficult to measure. 
No methods have yielded precise magnetic field strengths as far from 
the center as the virial radius; only lower limits using limits on 
IC emission have been placed. Equipartition arguments have been 
used \cite{2009A&A...506.1083V,2006Sci...314..791B,2006AJ....131.2900C,2009A&A...494..429B} but 
this gives only a rough estimate for the magnetic field strength and it 
relies on various assumptions that cannot be verified.  
The value of 5--7~$\mu$Gauss we find shows that a substantial magnetic 
field exists even far out from the cluster center.

Because radio relics directly pinpoint the location of shock fronts 
they can be used to get a complete inventory of shocks and their 
associated properties in galaxy clusters, important to understand 
the impact of shocks and mergers on the general evolution of clusters. 
Because less energetic mergers are more common and have lower Mach numbers, 
there should be many fainter relics with steep spectra which have currently escaped detection.   
Interestingly, these large-scale shocks 
in galaxy clusters have been suggested as acceleration sites for highly 
relativistic cosmic rays \cite{2003ApJ...593..599R}. As the radiation 
losses for relativistic cosmic ray protons are negligible, the maximum 
energy to which they can be accelerated is only limited by the lifetime 
of the shock, which can last for at least $10^{9}$ yr.  This means that 
in merging clusters protons can be accelerated up to extreme energies 
of $10^{19}$~eV, much higher than that in supernova remnants.

\clearpage

 \begin{figure}[f!]
    \begin{center}
      \includegraphics[angle = 90, trim =0cm 0cm 0cm 0cm,width=1.0\textwidth]{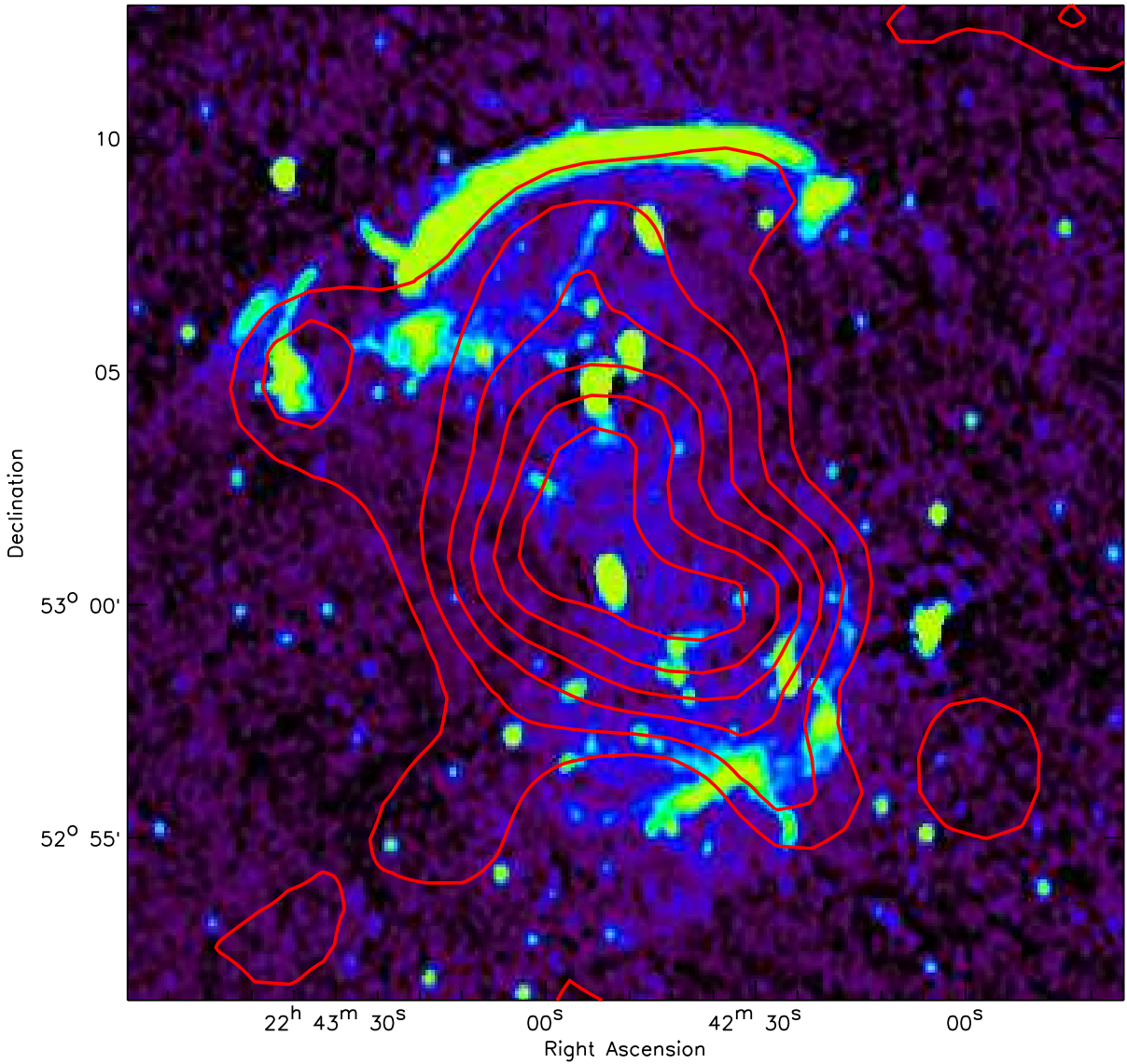}
       \end{center}
       \caption{WSRT radio image at 1.4~GHz. 
 The image has a resolution of 16.5~arcsec $\times$ 12.9~arcsec and the rms
 noise is 19~$\mu$Jy~beam$^{-1}$. Red contours (linearly spaced) represent the X-ray 
 emission from ROSAT showing the hot ICM.}
 \end{figure}

\begin{figure}[f!]
    \begin{center}
      \includegraphics[angle = 90, trim =0cm 0cm 0cm 0cm,width=1.0\textwidth]{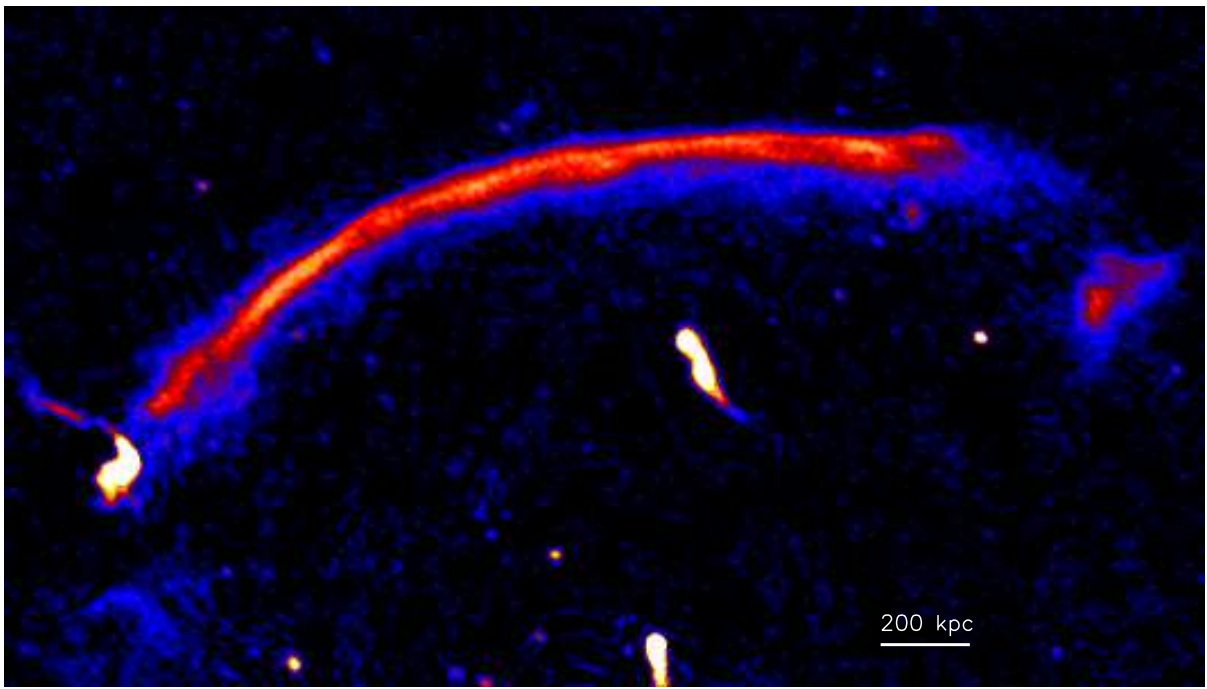}
       \end{center}
       \caption{GMRT 610~MHz radio image. 
 The image has a rms noise of 23~$\mu$Jy~beam$^{-1}$ and a  
 resolution of 4.8~arcsec $\times$ 3.9~arcsec.}
 \end{figure}

\begin{figure}[f!]
    \begin{center}
      \includegraphics[angle = 90, trim =0cm 0cm 0cm 0cm,width=0.7\textwidth]{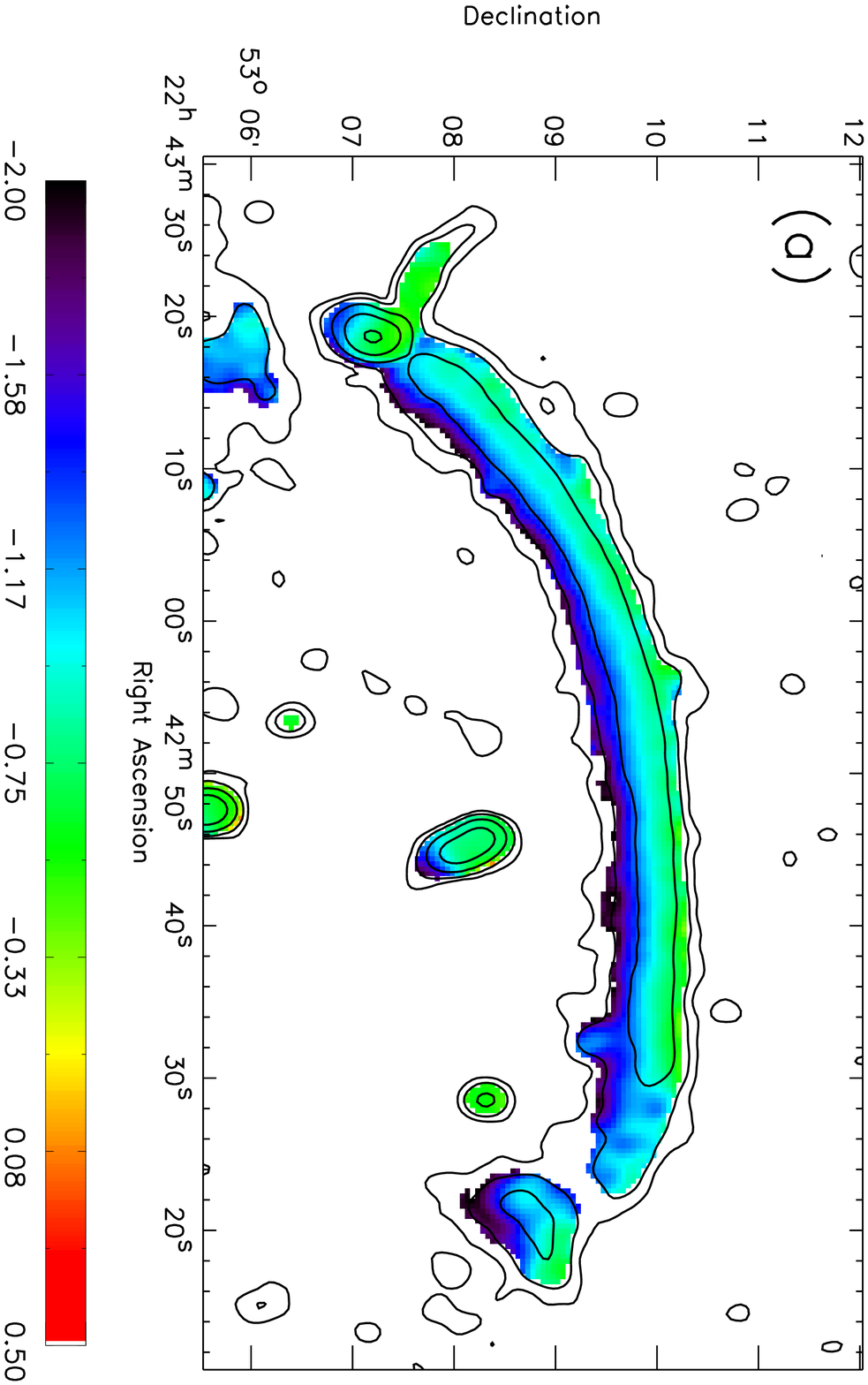}
      \includegraphics[angle = 90, trim =0cm 0cm 0cm 0cm,width=0.7\textwidth]{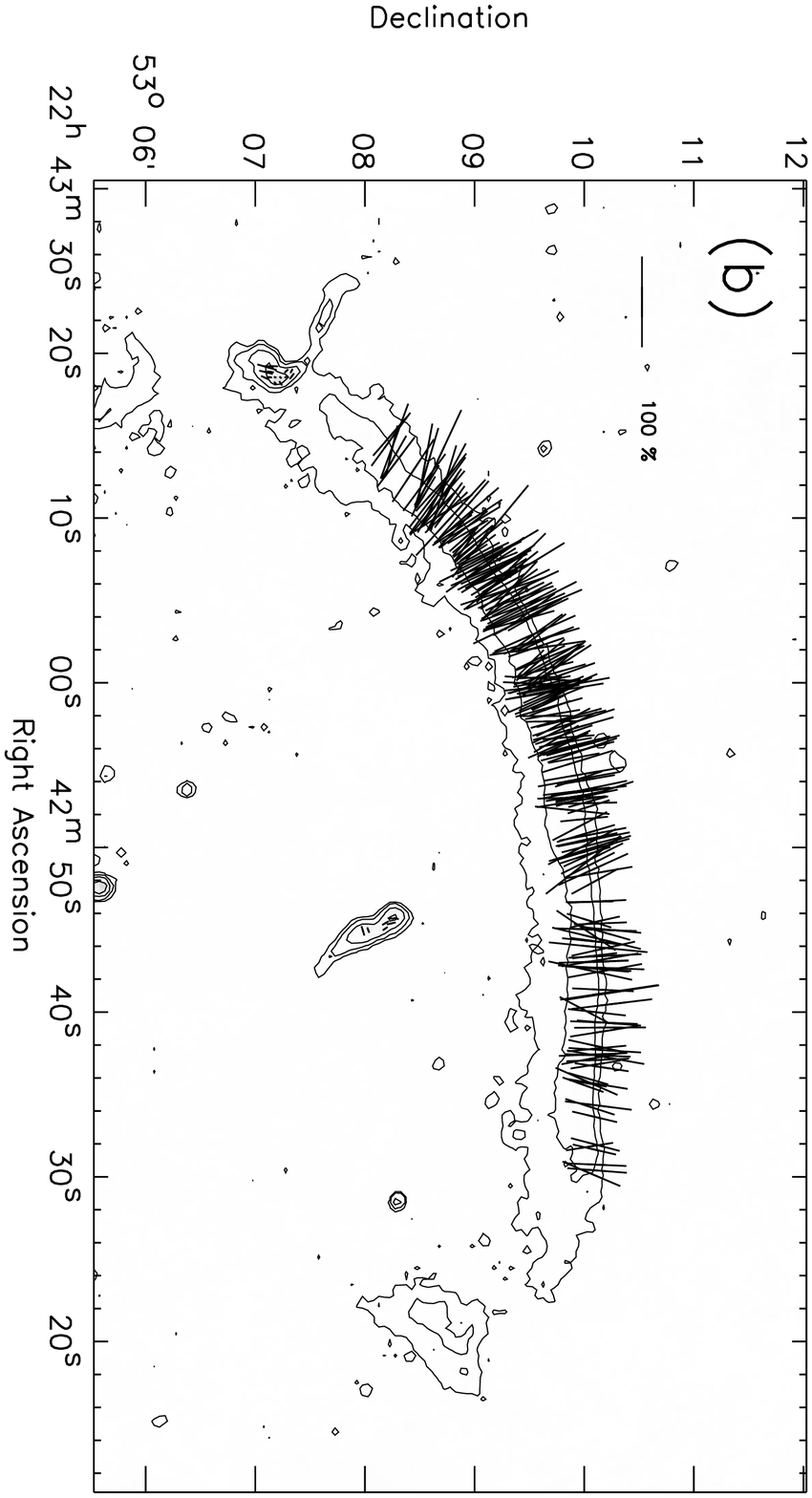}
       \end{center}
       \caption{Radio spectral index and polarization maps. {\it a}: 
 The spectral index was determined using matched observations at 2.3, 1.7, 1.4, 1.2, and 0.61~GHz, 
 fitting a power-law radio spectrum to the flux density measurements. 
 The map has a resolution of 16.7~arcsec $\times$ 12.7~arcsec.
 Contours are from the WSRT 1.4~GHz image and are drawn at levels 
 of $[1,4,16,\ldots] \times 36$~$\mu$Jy~beam$^{-1}$. 
  {\it b}:  The polarization 
 electric field vector map
 was obtained with the VLA at a frequency of 4.9~GHz and 
 has a resolution of 5.2~arcsec $\times$ 5.1~arcsec. 
 The contours are from Fig.~2 and are drawn at levels 
 of $[1,4,16,\ldots] \times 70$~$\mu$Jy~beam$^{-1}$. The length of the vectors 
 is proportional the polarization fraction, which is the ratio 
 between the total intensity and total polarized intensity. 
 A reference vector for 100\% polarisation is drawn in the top left corner. 
 The vectors were corrected for the effects of Faraday rotation 
 using a Faraday depth of $-140$~rad~m$^{-2}$ determined 
from the WSRT $1.2-1.8$~GHz observations.}
 \end{figure}

\begin{figure}[f!]
    \begin{center}
      \includegraphics[angle = 90, trim =0cm 0cm 0cm 0cm,width=1.0\textwidth]{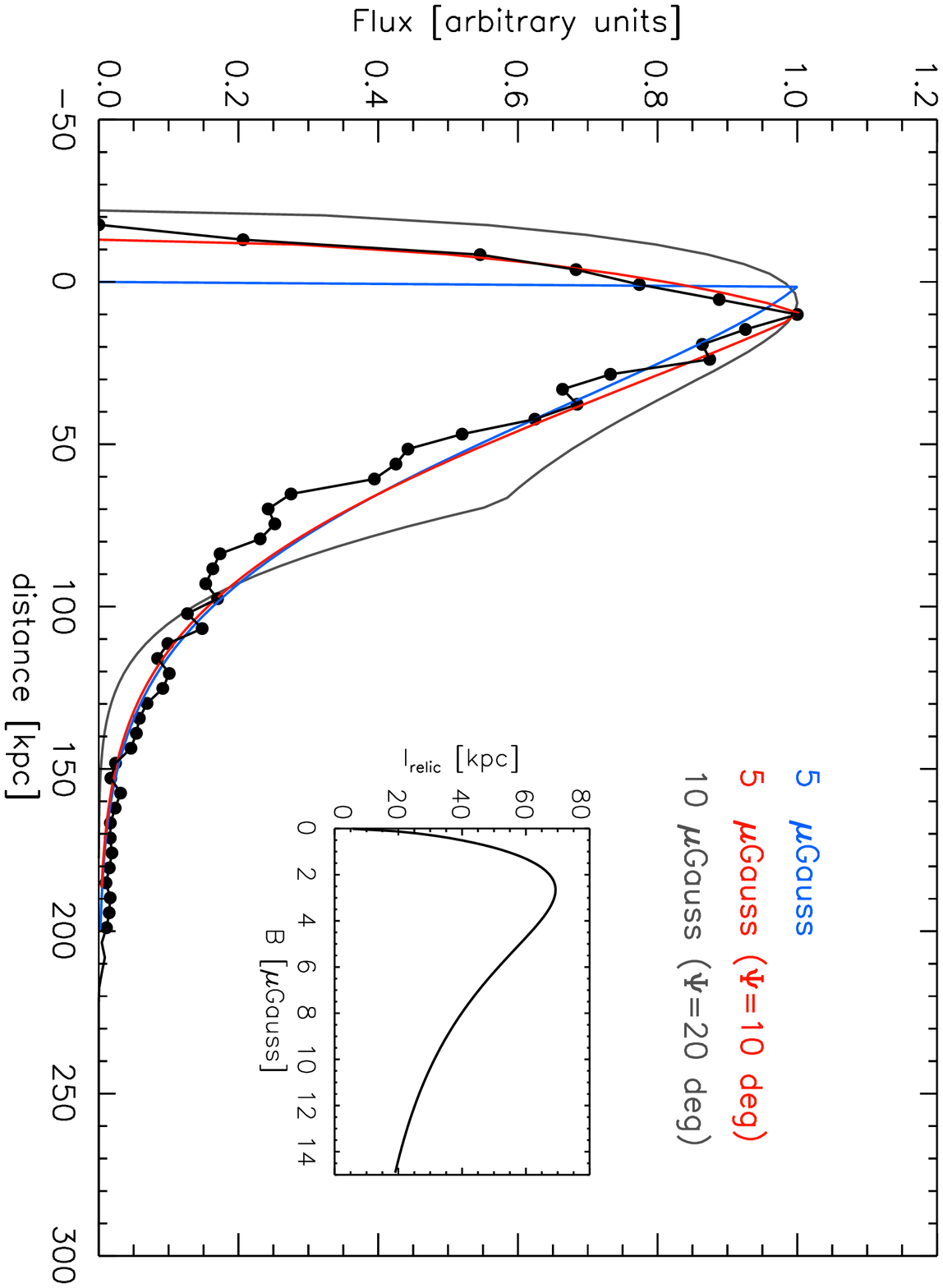}
       \end{center}
       \caption{The deconvolved profile at 610~MHz, averaged over the full length of the relic, 
 is shown by the solid black line and dots. 
 Models for different magnetic field strengths and projection 
 angles ($\Psi$; i.e., the angle subtended by the relic into the plane of the sky) are overlaid. 
 We used an equivalent magnetic field strength of the CMB at $z=0.1921$ of 4.6~$\mu$Gauss and a 
 downstream velocity of 1,000~km~s$^{-1}$. The model (red) 
 for $\Psi = 10$~deg and $B = 5$~$\mu$Gauss provides the best fit. 
 A model for $B = 5$~$\mu$Gauss without any projection effects is overlaid 
 in blue. For $\Psi > 15$~deg no good fit to the data could be obtained, 
 as an example we have plotted the profile (grey) for $\Psi = 20$~deg and $B = 10$~$\mu$Gauss.   
 (inset) The intrinsic width of the relic as function 
 of magnetic field strength (Eq.~1), it shows that for a given width usually 
 two solutions for the magnetic field strength can be obtained.}
 \end{figure}

\clearpage\newpage

\section*{Supporting Online Material (SOM) \newline \vspace{-5mm}\newline Materials and Methods}
We carried out radio observations of CIZA~J2242.8+5301 with the WSRT in 
the L-band (at 1.2, 1.4 and 1.7~GHz) and at 2.3~GHz recording full 
polarization products. The total integration time was 12~hr at 2.3, 
1.2 and 1.7~GHz, and 30~hr at 1.4~GHz. The observations were spread 
out over various runs between March and November 2009. For each different 
frequency setup the total bandwidth was 160~MHz. The 160~MHz was further 
divided over 8 sub-bands (IFs) with 20~MHz bandwidth and 64 spectral 
channels. GMRT observations at 610~MHz were carried out with 32~MHz 
bandwidth in spectral line mode with 512 channels on November 19, 2009. 
The total integration time was 9~hr. Only RR and LL polarization, 
to create a total intensity image, were recorded. VLA C-array 4.9~GHz 
observations were taken on August 17, 2009 in single channel continuum 
mode recording all four polarization products. Total integration time 
was about 7.5~hr.

We reduced the data using AIPS\footnote{http://www.aips.nrao.edu/} 
(Astronomical Image Processing System, NRAO) and CASA\footnote{http://casa.nrao.edu/} 
(Common Astronomy Software Applications). After inspection for 
the presence of radio frequency interference and other problems, bad data  
was subsequently removed (i.e., ``flagged''). Bandpass\footnote{only for the GMRT 
\& WSRT observations} and gain calibration were carried 
out using several bright unresolved calibrator sources. 
The flux scale was set using standard primary calibrators \cite{1977A&A....61...99B,perleyandtaylor}. 
For the WSRT observations 
the channel dependent polarization leakage terms were determined 
using a bright unpolarized calibrator source. The polarization angles were set 
using the polarized calibrators 3C138 and 3C286 for the WSRT and VLA observations. 
Subsequent rounds of self-calibration were carried out to improve the dynamic 
range of the images. Several bright nearby sources still limited the dynamic 
range in the 610~MHz GMRT image. These sources were subtracted from the data 
using the ``peeling-method'' \cite{2004SPIE.5489..817N,2009A&A...501.1185I}. The 610~MHz 
high-resolution image of the relic was made using robust 
weighting \cite{briggs_phd} set to $-1.0$. A 610~MHz image of the cluster, 
with robust weighting set to 0.5, is shown in Fig.~S5.

We made a radio spectral index map using images at 2.3, 1.7, 1.4, 1.2, 
and 0.61~GHz, fitting a power-law spectral index through the flux measurements. 
We limited the UV-ranges (to include only common UV-coverage) for the images that were
used to compute the spectral index map.
The spectral index map for the full cluster is shown in Fig.~S6. 
Both the northern and southern relics show steepening of the spectral 
index towards the cluster center.

We used the technique of Rotation Measure 
Synthesis \cite{2005A&A...441.1217B} 
to determine the Faraday depth of the northern relic. 
We found an average Faraday depth of about $-140$ rad~m$^{-2}$ and 
used that to correct for the effect of Faraday Rotation 
by de-rotating the electric/magnetic field vectors.

\renewcommand{\thefigure}{S\arabic{figure}}
 \begin{figure} [f!]
    \begin{center}
      \includegraphics[angle = 90, trim =0cm 0cm 0cm 0cm,width=1.0\textwidth]{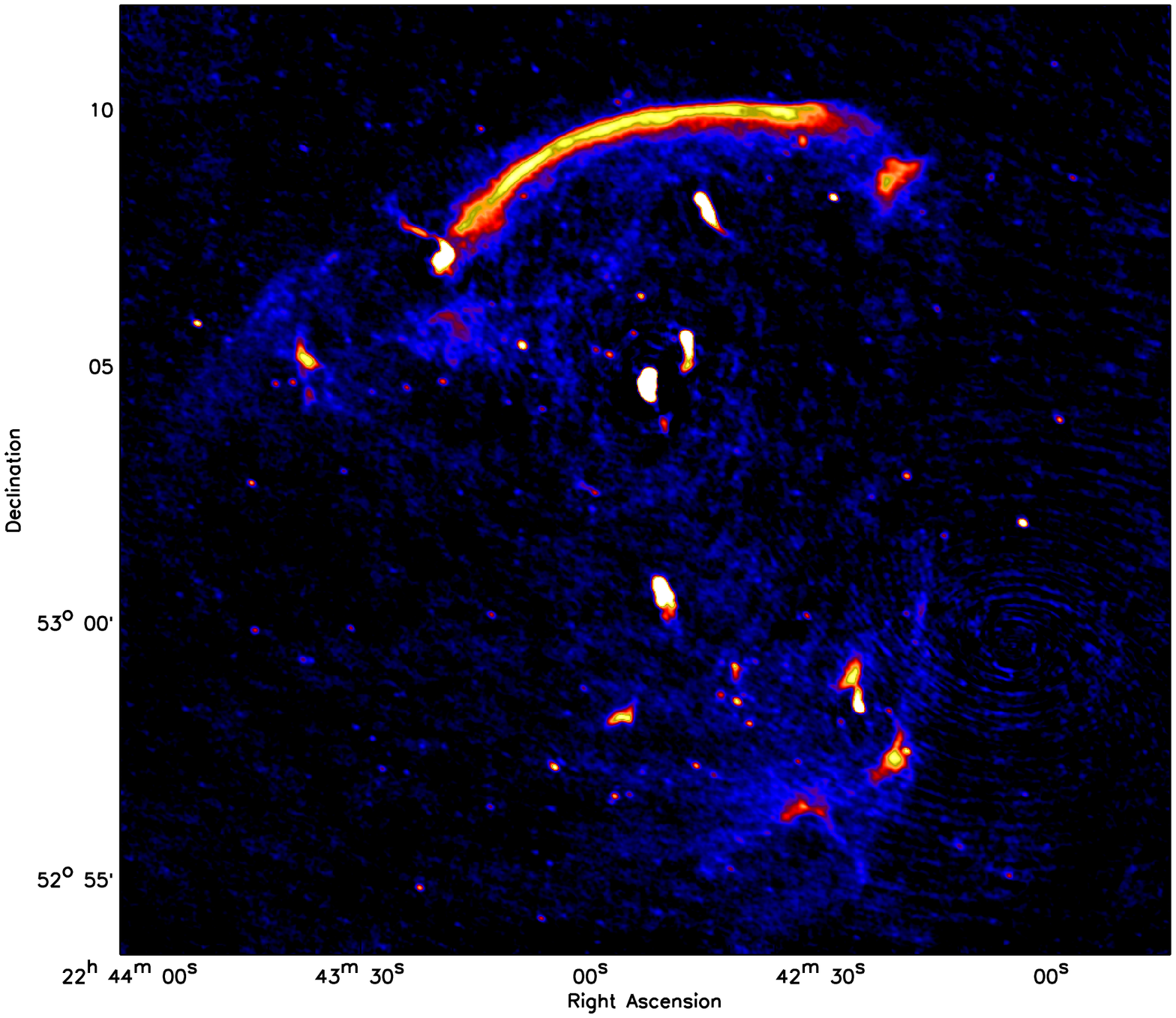}
       \end{center}
      \caption{GMRT 610~MHz radio image with a resolution of 
               5.8~arcsec~$\times$~4.4~arcsec. The rms noise in 
               the image is 24~$\mu$Jy~beam$^{-1}$.}
 \end{figure} 
 \begin{figure}[f!]
    \begin{center}
      \includegraphics[angle = 90, trim =0cm 0cm 0cm 0cm,width=1.0\textwidth]{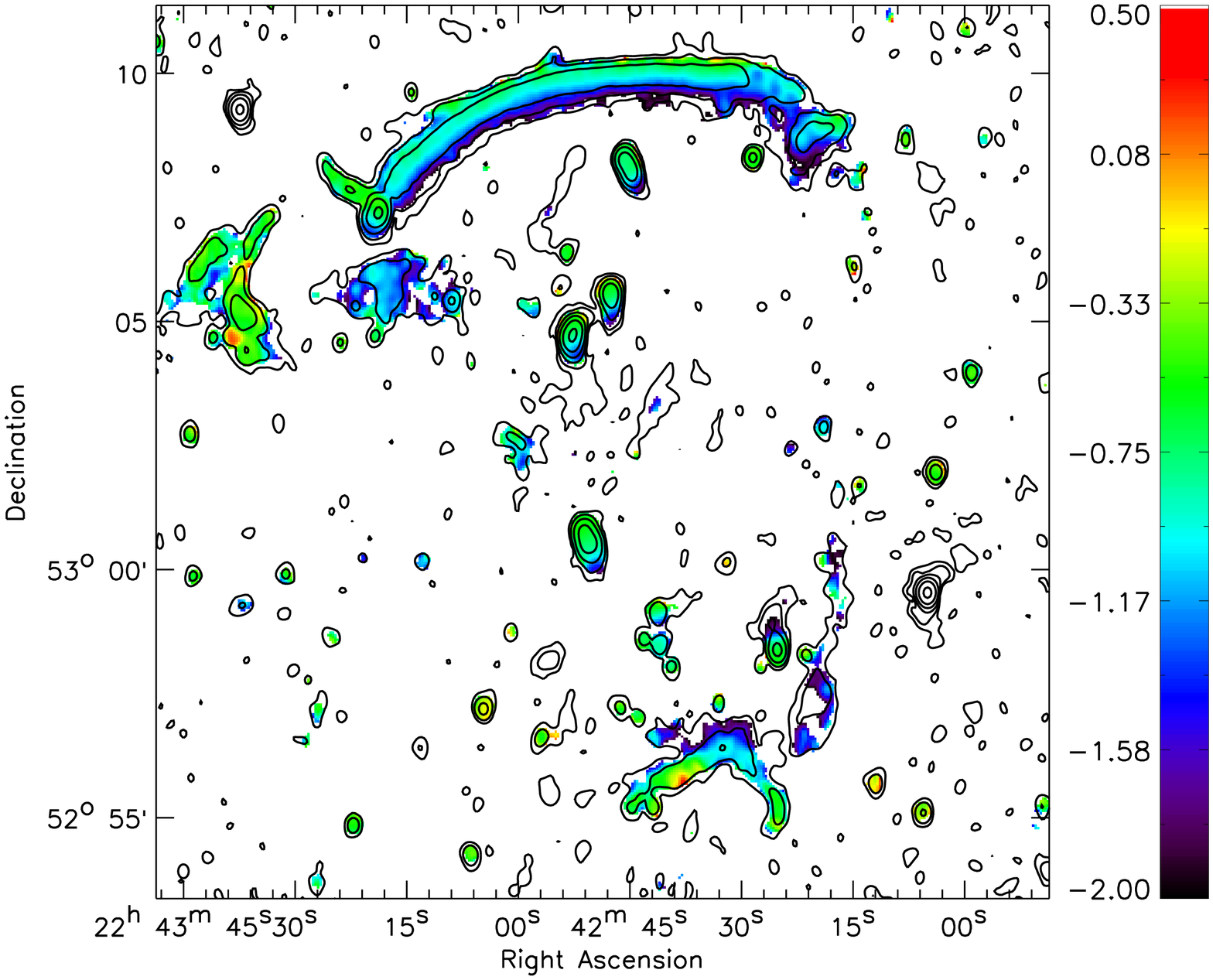}
       \end{center}
       \caption{Radio spectral index map. The spectral index 
                map has a resolution of 16.7~arcsec~$\times$~12.7~arcsec.
                Contours are from the WSRT 1.4~GHz image and are drawn at levels 
                of $[1,4,16,\ldots] \times 57$~$\mu$Jy~beam$^{-1}$. 
       }
 \end{figure}

\bibliography{1194293}

\bibliographystyle{Science}

\begin{scilastnote}

\item The WSRT is operated by ASTRON (Netherlands Institute for Radio Astronomy) 
with support from the Netherlands Foundation for Scientific Research (NWO). 
We thank the staff of the GMRT who have made these observations possible. 
The GMRT is run by the National Centre for Radio Astrophysics of the Tata Institute of Fundamental Research. 
The National Radio Astronomy Observatory is a facility of the National Science 
Foundation operated under cooperative agreement by Associated Universities, Inc. 
R.J.v.W. acknowledges funding from the Royal Academy of Arts and Sciences (KNAW). 
The authors thank G. Brunetti for discussions.

\end{scilastnote}
\end{document}